\documentclass{article}

\usepackage{amsfonts, amsmath, authblk, booktabs, float, graphicx, natbib}
\usepackage[margin=1in]{geometry}
\usepackage[T1]{fontenc}  

\title{Ball path curvature and in-game free throw shooting proficiency in the National Basketball Association}
\author[1]{Ruoqian Zhu}
\author[2]{Dave Love}
\author[3]{Scott Powers}
\affil[1]{Department of Mathematics, Rice University}
\affil[2]{NBA Shooting Coach}
\affil[3]{Department of Sport Management, Rice University}

\begin{document}
  \maketitle

  \begin{abstract}
    Basketball shooting coaches agree that smoother shooting motions are better, but there is less agreement about what ``smooth'' means quantitatively or what part of the shooting motion needs to be smooth. Using ball tracking data from the 2023-2024 National Basketball Association regular season, we explore the relationship between ball path curvature and free throw shooting performance. We fit Bézier curves to the ball tracking data in the sagittal plane and test different methods of calculating path curvature. We find that both max curvature and terminal curvature are negatively associated with shooting performance, but terminal curvature explains much more of the between-player variance in free throw shooting performance. This suggests that shooting coaches would be better off focusing on the smoothness at the end of the shot rather than at the beginning of the forward motion of the ball.
  \end{abstract}

  \section{Introduction}

    Free throws play a critical role in determining a basketball team's success, making the study of shooting mechanics an essential aspect of basketball performance analysis \citep{kozar_importance_1994}. Extensive research has been conducted to identify factors influencing free throw accuracy and to develop strategies for improvement \citep{tran_optimal_2008}. Within the realm of sports analytics, many hypotheses are rooted in conventional wisdom. Conventional coaching wisdom holds that smoother shots are better \citep{haefner_7_2010, penny_overlooked_2016}. However, smoothness has traditionally been assessed through subjective observation rather than a quantifiable measure, leaving room for inaccuracy and bias. Shooting coaches often focus on smoothing out the highest-curvature part of the smoothing motion, which is typically when the ball begins its final movement forward. But is this the right place to focus?
    
    \citet{slegers_role_2024} proposed a framework for analyzing the path of the ball during the shooting motion using curvature-based metrics and collected data from jump shots attempted by male professional basketball players in a controlled, experimental environment. Surprisingly, they found max ball path curvature to be positively associated with shooting performance. Rather, terminal ball path curvature (more weight on curvature closer to release) was negatively associated with shooting performance. In other words, shooters should not want to smooth out the peak curvature---they should want to smooth out the curvature as the ball approaches release. The recommended focus on terminal path curvature makes intuitive sense because it means keeping the path as smooth as possible when the ball is moving the fastest.

    The present study builds directly on the methodology of \citet{slegers_role_2024}, making adaptations to handle in-game data from the National Basketball Association (NBA) collected via Second Spectrum. This adaptation involves pre-processing noisy ball-tracking data, implementing ball-path standardization methods for fair comparison across players, and regressing the metrics directly against season free throw shooting percentage (FT\%). In doing so, this study not only seeks to validate the findings of \citet{slegers_role_2024} in game conditions, but also contributes a robust and generalizable framework for evaluating shot smoothness and shooting accuracy at the elite level.

  \section{Related work}
  
    The analysis of basketball shooting mechanics has been an ongoing research subject in sports analytics, with free throw performance being a specialized area of study. Despite the unique and uncontested nature of free throws with their standardized execution, many of the biomechanical principles and methodologies used to investigate general shooting mechanics apply to free throws as well. Traditional research has primarily focused on the kinematic and kinetic factors that influence shooting performance, such as body alignment, segmental coordination, force generation, and energy transfer. This section reviews key contributions to the field, with a particular focus on ball trajectory and release conditions.

    \subsection{Individualized Optimization in Shooting Mechanics}

      Early research on optimizing basketball shooting mechanics sought to determine the optimal configuration of key factors, including release height, release speed, launch angle, side angle, and backspin, by analyzing hundreds of thousands of simulated basketball trajectories \citep{tran_optimal_2008}. Findings suggest that each of these factors has an optimal threshold that shooters should aim to achieve. For instance, it is recommended that shooters apply up to 3 Hz of backspin, launch the ball at an angle of 52 degrees relative to the horizontal, and release the ball from the highest possible position \citep{tran_optimal_2008}.

      However, more recent research has shown that optimal shooting mechanics, specifically release conditions, must be highly individualized to achieve the best result. Using surface electromyography and high-speed cameras, \citet{pakosz_muscle_2021} revealed that elite basketball players manifest increased variability in muscular activation time relative to their less skilled counterparts. This variability is likely due to extensive training that allows players to fine-tune their muscle activation timing with specific shot requirements. This study concluded that there is no universal way to artificially increase or decrease muscle activation time, and optimal shooting performance is highly dependent on one's ability to control and adjust muscle activation in an individualized manner. 

      A similar concept of individualized optimization in shooting mechanics has been applied to release strategies, including release angle and velocity. Research has shown that maximum shooting performance is achieved when athletes match their release parameters to their unique mechanics, rather than conforming to a predetermined ideal release angle \citep{slegers_basketball_2022}. These findings reinforce the idea that an optimal shot is not universal but highly specific to players, shaped by their experiences, physical attributes, and muscle coordination. 

    \subsection{Ball Path Curvature and Shooting Accuracy}
    
      In addition to release conditions, recent research has emphasized the role of ball trajectory and curvature in determining shooting accuracy. \citet{slegers_role_2024} revealed a negative correlation between the terminal curvature of the ball path and longitudinal shooting accuracy, with the latter being assessed using the standard deviation of the intra-individual release velocity as a proxy, given its relevance to distance control. Furthermore, \citet{slegers_role_2024} validated that in the S-shaped shooting path, better shooters exhibit a sharper transition from the backswing to the forward shooting motion compared to less skilled players. Their study involved 31 professional male basketball players, each of whom attempted 40 jump shots under controlled conditions. 

    \subsection{Smoothness Quantification and Other Release Factors}

      The principle that ``smoother'' shots are better is commonly emphasized in coaching resources, where smooth, continuous release is considered essential to achieving repeatable, high-percentage shooting mechanics \citep{haefner_7_2010, penny_overlooked_2016}. The concept of smoothness has also been mentioned in \citet{pakosz_muscle_2021} and \citet{tran_optimal_2008}, both emphasizing that it is desirable for the players' shooting movement to be as smooth as possible. However, neither study provides a quantifiable measure of smoothness, highlighting a gap in the literature for the development of scientific metrics. More granular analysis of the ball’s trajectory is necessary to gain a deeper understanding of shooting mechanics.

      Previous research has shown that the angular motions of the elbow and wrist joints compensate for each other toward the end of the shooting motion to adjust for subtle variations in the ball release parameters \citep{button_examining_2003}. Similarly, \citet{tran_optimal_2008} focused solely on the release phase of the shot. This finding highlights the important role of the latter stages of the shot, which implies that the influence of the ball path is not uniform throughout the trajectory. This insight presents opportunities for more detailed research into how different stages of the trajectory contribute to shooting accuracy and consistency.

      Velocity, a fundamental factor in any motion analysis, has been a subject of debate regarding its role in impacting shooting accuracy. \citet{pakosz_muscle_2021} suggested that free throw speed does not play a significant role in determining accuracy. In contrast, \citet{hamilton_optimal_1997} found that an optimal trajectory should have an initial release angle of approximately 60 degrees and a speed of 7.3 m/s. This particular combination is advantageous as it directs the ball closer to the far rim of the basket rather than the near rim. \citet{slegers_role_2024} suggested that the ability to control the release speed is important, and variations in the final ball path could increase the difficulty of speed control, potentially reducing shot accuracy. \citet{mullineaux_coordination-variability_2010} acknowledged that an optimal minimal release speed exists, and successful shots are associated with release speeds closer to the optimal minimum speed than failed shots. Moreover, increased elbow-wrist coordination variability right before release was observed in failed shots, indicating last-moment adjustments due to perceived errors. These findings demonstrate that the coordination between elbow and wrist angles plays a crucial role in generating the appropriate ball speed.
      
  \section{Data}

    Our dataset comprises 515 free throws attempted in games during the 2023-2024 NBA regular season. These free throws were attempted by 35 distinct players, with each player contributing approximately 15 free throw attempts. We obtained optical tracking data from the Second Spectrum system \citep{national_basketball_association_nba_2016}, which captured the $(x, y, z)$ position of the basketball at 25 frames per second throughout each shot's trajectory. This granularity enables a detailed analysis of ball motion dynamics, facilitating the development of curvature-based metrics to quantify shooting smoothness. For each of the 35 players, we obtained their 2023-2024 season free throw shooting percentage (FT\%) from \citet{basketball_reference_2023-24_2024}.

\begin{figure}[H]
        \centering
        \includegraphics[scale=0.3]{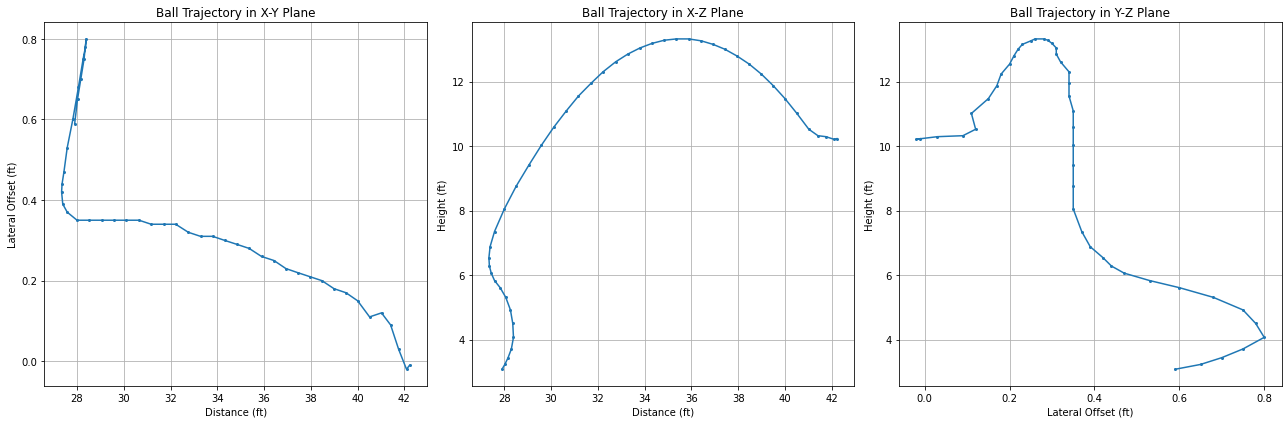}
        \caption{\it Three-view projection of a single free throw trajectory captured using ball tracking data. The panels display the ball's motion in the X-Y plane (view from behind shooter), X-Z plane (view from the side) and Y-Z plane (view from above), respectively. The trajectory begins at the start of the shooting motion and ends when the ball arrives at the rim. Note the compressed scale in the Y dimension (lateral offset).}
        \label{fig:data-description}
      \end{figure}

    Figure \ref{fig:data-description} illustrates an example of a free throw trajectory measured by the optical tracking system, projected onto the three principal planes. The side view (X-Z plane) illustrates the characteristic S-shaped shooting curve, followed by the arc of the ball as it approaches the basket. In the sections that follow, we quantify the smoothness of the shot using the path traced by the basketball during the shooting motion. Because we do not observe the movement of the players' limbs, we rely on the movement of the basketball to describe the smoothness of the shot.

  \section{Methods}
  
    The methods in this study build upon the curvature-based modeling approach developed by \citet{slegers_role_2024}. We adapt and extend their methods to handle in-game tracking data. Specifically, this section outlines how curvature and directional metrics are computed from noisy spatiotemporal data, how ball trajectories are standardized across players, and how the relationships between the resulting metrics and the free throw shooting performance are analyzed. These adaptations are critical for testing whether the negative association between late-path curvature and shooting accuracy observed in experimental settings holds true under high-pressure, in-game conditions, and for establishing a more generalizable, automated framework for evaluating free throw mechanics at the elite level.

    \subsection{Path Standardization}

\begin{figure}[H]
        \centering
        \includegraphics[scale=0.4]{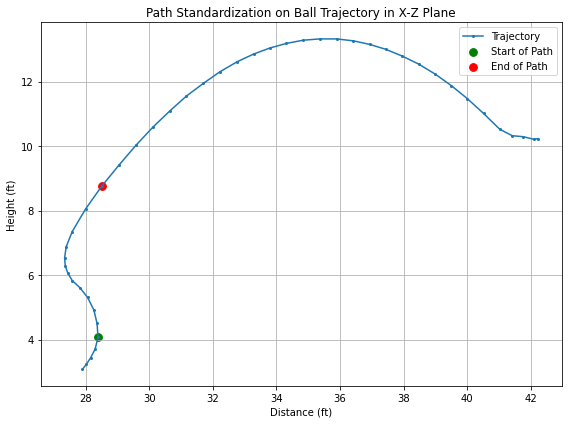}
        \caption{\it Ball trajectory in the X-Z plane for a single free throw attempt. The green marker indicates the start of the analyzed trajectory, defined as the frame where the ball is furthest from the player’s body. The red marker indicates the end point, defined as two frames after the ball reaches its maximum velocity, approximating the moment of release.}
        \label{fig:path-standardization}
      \end{figure}

      The first step in analyzing the path of the ball during the shooting motion is to identify the start and the end of the path for each shot. Our goal is to identify the time frame from the beginning of the shooting motion to the release of the shot. \citet{slegers_role_2024} defined the start of the ball path as the point at which the ball rose above the shooter's shooting-hand elbow and defined the end of the ball path as the final frame in which the ball remained in contact with the shooter's hand. However, in our data we do not have any information about shooter's elbow or hand---only the ball. To address this limitation, we define the start of the ball path as the frame in which the ball is furthest from the player's body, corresponding to the first elbow of the S-shaped shooting curve. We define the release point as two frames (0.08 seconds) after the frame at which the ball reaches its maximum speed. This rule is based on the assumption that ball speed is maximized at release, erring on the side of including more frames in case of measurement error. Figure \ref{fig:path-standardization} illustrates the start and end points of the ball path for the free throw attempt introduced in Figure \ref{fig:data-description}.

      After completing path standardization, we are left with $m + 1$ observed data points \((x_0, y_0, z_0)\), \((x_1, y_1, z_1)\), ..., \((x_m, y_m, z_m)\) along the ball path for each free throw, evenly spaced in time. We parameterize time for each shot so that the first observation occurs at time $\tau = 0$ and the last observation occurs at time $\tau = 1$.

    \subsection{Curvature}
    \label{sec:curvature}
    
    To quantify shot smoothness and examine its relationship with free throw accuracy, this section introduces a curvature-based metric that captures the degree of bending in the ball's trajectory. Following \citet{slegers_role_2024}, we calculate curvature in the X-Z plane, ignoring lateral movement of the ball during the shooting motion. Bézier curves are particularly advantageous for this purpose due to their smoothness, continuity and differentiability, which allow for precise curvature computation. 

        \subsubsection{Bézier Curves}

            Bézier curves, widely utilized in computer-aided graphical design, are parametric curves defined by a set of unique control points. The coordinates of a Bézier curve are polynomial functions given by a set of control points weighted according to the coefficients of a Bernstein polynomial \citep{baydas_defining_2019}. As a result, Bézier curves are both compact and differentiable, making them well-suited for modeling smooth trajectories. These properties make Bézier curves an attractive approach for estimating the ball path and deriving curvature-based metrics \citep{slegers_role_2024}. A Bézier curve is defined by \(n + 1\) control points \(p_0 = (x_0^*, z_0^*), ..., p_n = (x_n^*, z_n^*)\). For any time \(\tau \in [0, 1]\), the location of the curve is given by:
            \begin{equation}
              \label{eqn:bezier}
              B_n(\tau) = \sum_{i=0}^{n} p_i \binom{n}{i} (1 - \tau)^{n-i} \tau^i,
            \end{equation}
            where \( \binom{n}{i} \) are binomial coefficients, i.e. $n! / i! (n - i)!$. The {\it order} of the Bézier curve is \(n\), and the curve is a weighted averaged of the \(n+1\) control points with weights that vary from start to finish. Note that the location of the curve is \(p_0\) at time \(\tau =0\) and \(p_n\) at time \(\tau=1\). The first and last control points lie on the curve, but the other \(n - 1\) control points generally do not lie on the curve.

            For an individual free throw, we seek a Bézier curve which closely approximates the discrete points observed along the ball's trajectory. From this curve, we can then calculate curvature. We use \(\mathbf{Y}\) to denote the \( (m + 1) \times 2\) matrix of observed points at evenly spaced time points $\tau_0 = 0,\, \tau_1 = 1 / m,\, ...,\, \tau_m = 1$ along the ball's trajectory. To find the best-fit curve for these points, it is helpful to express \(B_n(\tau)\) in matrix form:
            \[
              B_n(\tau) = \mathbf{T}_n(\tau) \times \mathbf{M}_n \times \mathbf{P}_n,
            \]
            where \( \mathbf{T}_n(\tau) \) is a \( 1 \times (n+1) \) row vector with elements \( \tau^0 \) to \( \tau^n \); \( \mathbf{P}_n \) is an \( (n+1) \times 2 \) matrix containing the \( n+1 \) control points in the form \( (x_i, y_i) \); and \( \mathbf{M}_n \) is the \( (n+1) \times (n+1) \) lower-triangular matrix given by:
            \[
              \mathbf{M}_n[i,\, j] = \begin{cases}
                  \binom{n}{i} \binom{n - i} {j - i} (-1)^{j - i} & \mbox{if } j \ge i\\
                  0 & \mbox{otherwise}
              \end{cases}.
            \]
            Then the $m$ points along the Bézier curve approximating the $m$ observed data points are given by $\mathbf{T}_n \times \mathbf{M}_n \times \mathbf{P}_n$, where \( \mathbf{T}_n \) is an \( (m + 1) \times (n+1) \) matrix with each row being \( \mathbf{T}_n(\tau) \) for $\tau = \tau_0, ..., \tau_m$.

            The least-squares fit is the one which minimizes the squared error between the observed data points and the approximating points on the Bézier curve:
            \[
              \hat{\mathbf{P}}_n = \arg\min_{\mathbf{P}_n \in \mathbb{R}^{n \times 2}} ||\mathbf{Y} - \mathbf{T}_n \mathbf{M}_n \mathbf{P}_n||.
            \]
            This optimization problem has a closed-form analytical solution:
            \[
              \hat{\mathbf{P}}_n = \mathbf{M}_n^{-1} (\mathbf{T}_n^\top \mathbf{T}_n)^{-1} \mathbf{T}_n^\top \mathbf{Y}.
            \]
            
\begin{figure}[H]
                \centering
                \includegraphics[scale=0.4]{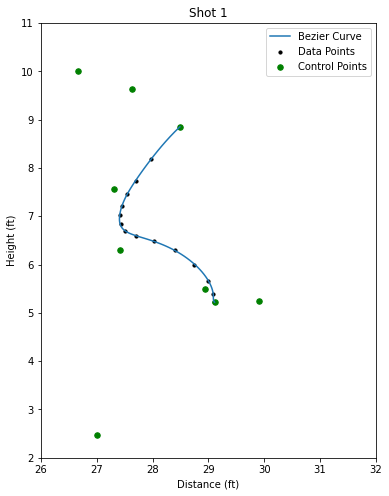}
                \caption{\it Free throw shot modeled by a Bézier curve with nine control points. The raw data points from the shot are shown in black, while the fitted curve is shown in blue. The green points represent the control points derived through least squares optimization.}
                \label{fig:free-throw-bezier}
            \end{figure}

            Figure \ref{fig:free-throw-bezier} shows an example of a free throw shot modeled using an eighth-order Bézier curve (i.e. nine control points). As seen in the figure, the curve closely follows the observed data, capturing both the smooth upward arc and the transition into the release phase of the shot. Throughout this study, we use $n = 8$ because we found this to yield satisfactory curve approximations via visual inspection of our data. This is similar to \citet{slegers_role_2024}, who used $n = 6$.
            
        \subsubsection{Curvature Calculation}

          In this section, to quantify the local smoothness of the ball’s trajectory, we compute the instantaneous curvature of the Bézier curve at each point along the shooting path. Curvature serves as a fundamental geometric descriptor of motion, enabling granular analysis of the fluctuations and smoothness of the shot over time. Curvature measures how sharply the shooting trajectory bends at a specific point. Following \citet{slegers_role_2024}, we use the following definition of curvature in two dimensions:
          \[
            \kappa(\tau) = \frac{||B_n'(\tau) \times B_n''(\tau)||}{||B_n'(\tau)||^3}.
          \]
          
          To calculate this curvature, we begin by analyzing the parametric derivatives of the Bézier curve. The first derivative of \(B_n(\tau)\) is obtained by differentiating the Bézier formulation (\ref{eqn:bezier}) and can be shown (with some simplification of terms) to be:
          \[ 
            B_n'(\tau) = n \sum_{i=0}^{n-1} (p_{i+1} - p_i) B_{i,n-1}(\tau),
          \]
          where \(B_{i, n}(\tau) = \binom{n}{i}(1 - \tau)^{n-i}\tau^i\) is the weight on $p_i$ in $B_n(\tau)$. Similarly, the second derivative \( B''(\tau) \) is
          \[
            B_n''(\tau) = n(n-1) \sum_{i=0}^{n-2} (p_{i+2} - 2p_{i+1} + p_i) B_{i,n-2}(\tau).
          \]
          Using these two equations, the curvature of \(B_n(\tau)\) = \((x(\tau), z(\tau))\) is given by:
          \begin{equation}
          \label{eqn:curvature}
            \kappa(\tau) = \frac{\left| x'(\tau) z''(\tau) - z'(\tau) x''(\tau) \right|}{\left( (x'(\tau))^2 + (z'(\tau))^2 \right)^{3/2}}.
          \end{equation}
    
        \subsubsection{Weighted Curvature Integral}
        \label{sec:weighted-curvature-integral}
    
        
          We hypothesize that ball path smoothness is correlated with shooting success, particularly as the ball approaches release (smoothness earlier in the ball path is less important). This section introduces the weighted curvature integral to summarize the smoothness of the full ball path with greater emphasize later in the path. While the curvature (\ref{eqn:curvature}) is a function of a point in time, the integral aggregates curvature values over the entire shooting motion. By adjusting the weighting scheme, this metric allows the analysis to emphasize the latter stages of the shot, which are most relevant to shot control and release \citep{button_examining_2003}.

          Following \citet{slegers_role_2024}, we introduce an integer weighting factor $\ell$ to calculate a weighted integral of the curvature with more weight as time increases:
          \begin{equation}
          \label{eqn:weighted-curvature-integral}
            \sigma(\ell) = \int_{0}^{1} (\ell + 1) \kappa(\tau)^\ell \, d\tau.
          \end{equation}
          If \(\ell = 0\), \(\sigma(\ell)\) is precisely the unweighted average curvature of the ball path. As $\ell$ increases, the integral (\ref{eqn:weighted-curvature-integral}) places more weight toward the latter part of the shot, converging to \(\kappa(1)\) as $\ell$ goes to infinity. 
    
\begin{figure}[H]
              \centering
              \includegraphics[width=0.8\linewidth]{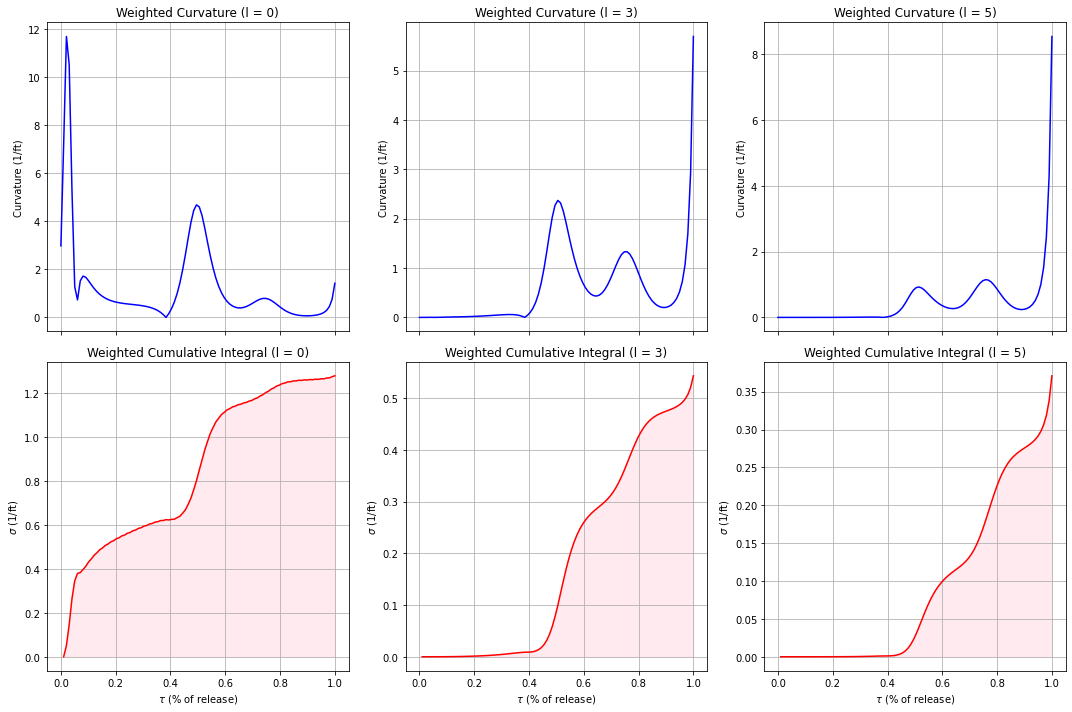}
              \caption{\it Weighted curvature and curvature integrals for the free throw shot from Figure \ref{fig:free-throw-bezier}. The top row shows the product of time-weight and curvature, and the bottom row shows the time-weighted curvature integral \(\sigma(\ell)\), for \(\ell = 0, 3, 5\). The top-left panel shows the unweighted curvature (\(\ell = 0\)).}
              \label{fig:weights}
          \end{figure}

          Figure \ref{fig:weights} visualizes how different values of $\ell$ affect the curvature weighting. The top row shows the curvature profiles at $\ell$ = 0, 3, and 5, while the bottom row displays their corresponding cumulative integrals. As $\ell$ increases, the curvature approaching the end of the ball path dominates the integral, leading to a more pronounced weighting of end-path behavior. This enables the metric to better capture the smoothness characteristics that may be more influential of shot success.

    \subsection{Weighted Least Squares Regression}

      The purpose of the curvature metrics described in the previous section is to quantify the smoothness of each shot. Ultimately, our goal is to test the association between shot smoothness and shooting accuracy. While \citet{slegers_role_2024} measured shooting performance using intra-individual release velocity standard deviation in a controlled setting, this metric is not readily available or reliable for in-game data due to the lack of biomechanical tracking data, which makes it difficult to precisely identify the moment of ball release. Therefore, we use player-level free throw percentage (FT\%) as the outcome variable. FT\% reflects real-world shooting performance under competitive, high pressure and is widely accepted as a summary statistic for evaluating a player’s free throw success over time. It comes with the added benefit that our sample size for estimating season-long FT\% is much greater than the approximately 15 shots with ball tracking data we observe for each player.

      To test the relationship between shot smoothness and FT\%, we calculate each player's average curvature across the free throws for which we have ball tracking data, and we regress season FT\% onto different definitions of average curvature via weighted least squares (WLS). Using \(\mbox{FTA}_p\), \(\mbox{FT\%}_p\) and \(C_p\) to denote player $p$'s season free throw attempts, season FT\%, and curvature metric (i.e. max curvature or weighted curvature integral), respectively, we estimate the following model:
      \begin{equation}
      \label{eqn:weighted-least-squares}
        \mbox{FT\%}_p \sim \mbox{Normal}(\alpha + \beta \cdot C_p, \sigma^2 / \mbox{FTA}_p),
      \end{equation}
      assuming independence between different players' outcomes. In effect, this weights each observation by the number of free throws attempted by the player during the full season. This means that players who shot more free throws in the 2023-2024 season, and hence have more precise shooting performance estimates, exert greater influence on the model. For each curvature metric, we obtain a $p$-value by testing the null hypothesis $\mathcal{H}_0: \beta = 0$ against the two-sided alternative $\mathcal{H}_A: \beta \ne 0$.
    
\section{Results}
    
    The metrics computed for this study include (a) max curvature $\kappa_{\max} = \max_\tau[\kappa(\tau)]$ and (b) time-weighted curvature integrals with weighting factors $\ell$ = 0, 3, and 5. For each player, these metrics are computed on a per-shot basis and subsequently averaged across all of their recorded free throw attempts (approximately 15 shots per player) to obtain player-level estimates.

\begin{figure}[H]
        \centering
        \includegraphics[width=0.8\linewidth]{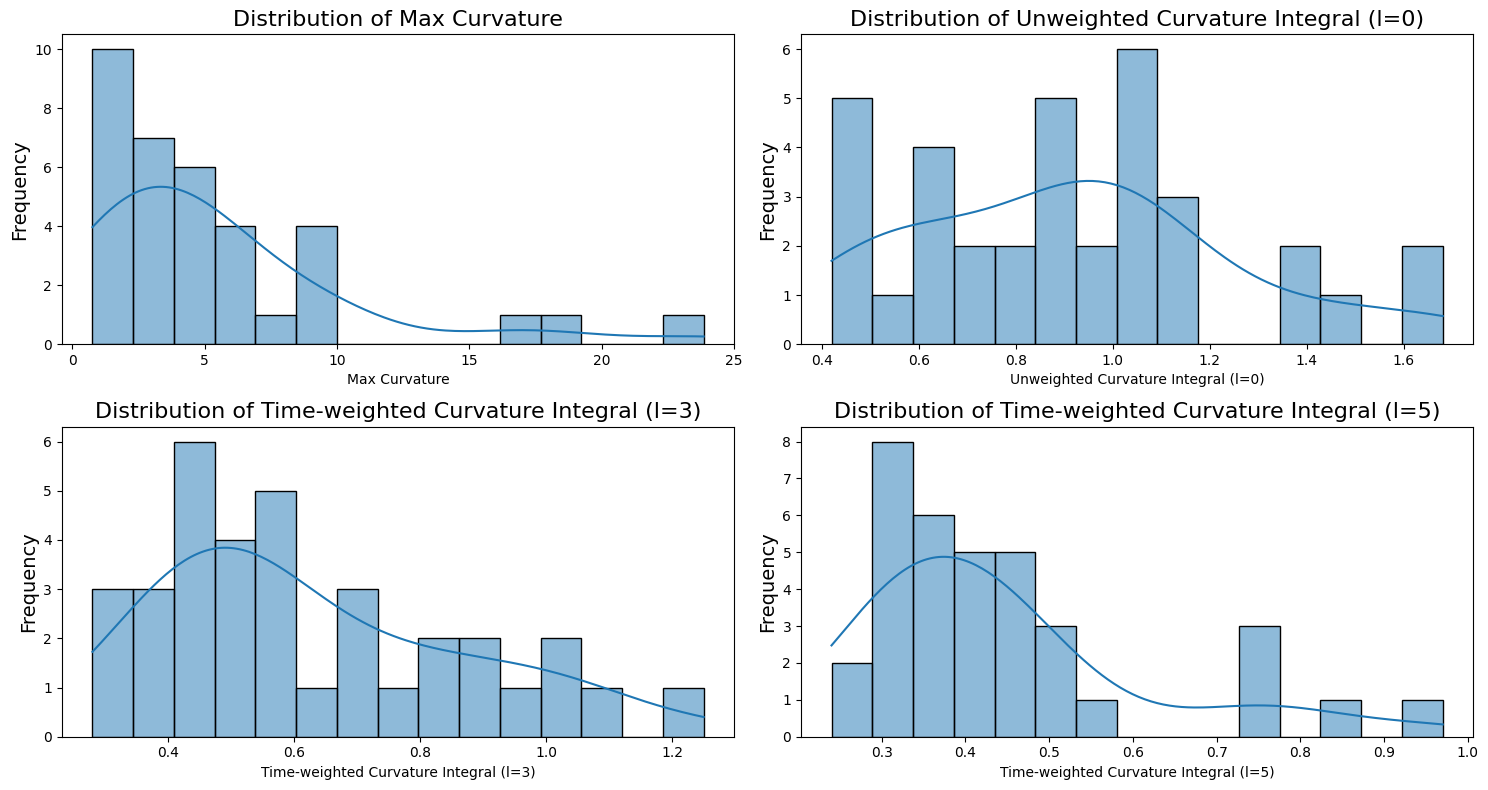}
        \caption{\it Distributions of key shot smoothness metrics across all players in the study. Kernel density estimate (KDE) curves are used to illustrate the underlying shape of each distribution.}
        \label{fig:scattered}
    \end{figure}

    Figure \ref{fig:scattered} shows the distribution of FT\% and all curvature metrics among the 35 players in the sample. FT\% appears to be approximately symmetric but slightly skewed to the left, which is consistent with most NBA players achieving moderate to high accuracy on the free throw line. All of the curvature metrics are right-skewed, reflecting that a few players have large peak or terminal curvatures.

    Table \ref{tab:summary-stats} provides summary statistics for each metric. The range of FT\% spans from 50.6\% to 92.7\%, with a standard deviation of 12.4\%. Notably, max curvature demonstrates considerable variability (SD = 5.15), while the curvature integrals are more tightly distributed.

    \begin{table}[H]
        \centering
        \begin{tabular}{lrrc}
        \toprule
        \textbf{Metric} & \textbf{Mean} & \textbf{Std Dev} & \textbf{Range} \\
        \midrule
        FT\% & 74.29 & 12.39 & 50.60--92.70 \\
        Max Curvature & 5.50 & 5.15 & 0.76--23.85 \\
        Unweighted Curvature Integral ($\ell=0$) & 0.91 & 0.34 & 0.42--1.68 \\
        Time-weighted Curvature Integral ($\ell=3$) & 0.63 & 0.25 & 0.28--1.25 \\
        Time-weighted Curvature Integral ($\ell=5$) & 0.45 & 0.17 & 0.24--0.97 \\
        \bottomrule
        \end{tabular}
        \caption{\it Descriptive statistics of free throw percentage and ball path curvature metrics ($n = 35$).}
        \label{tab:summary-stats}
    \end{table}

\begin{figure}[H]
        \centering
        \includegraphics[width=0.6\linewidth]{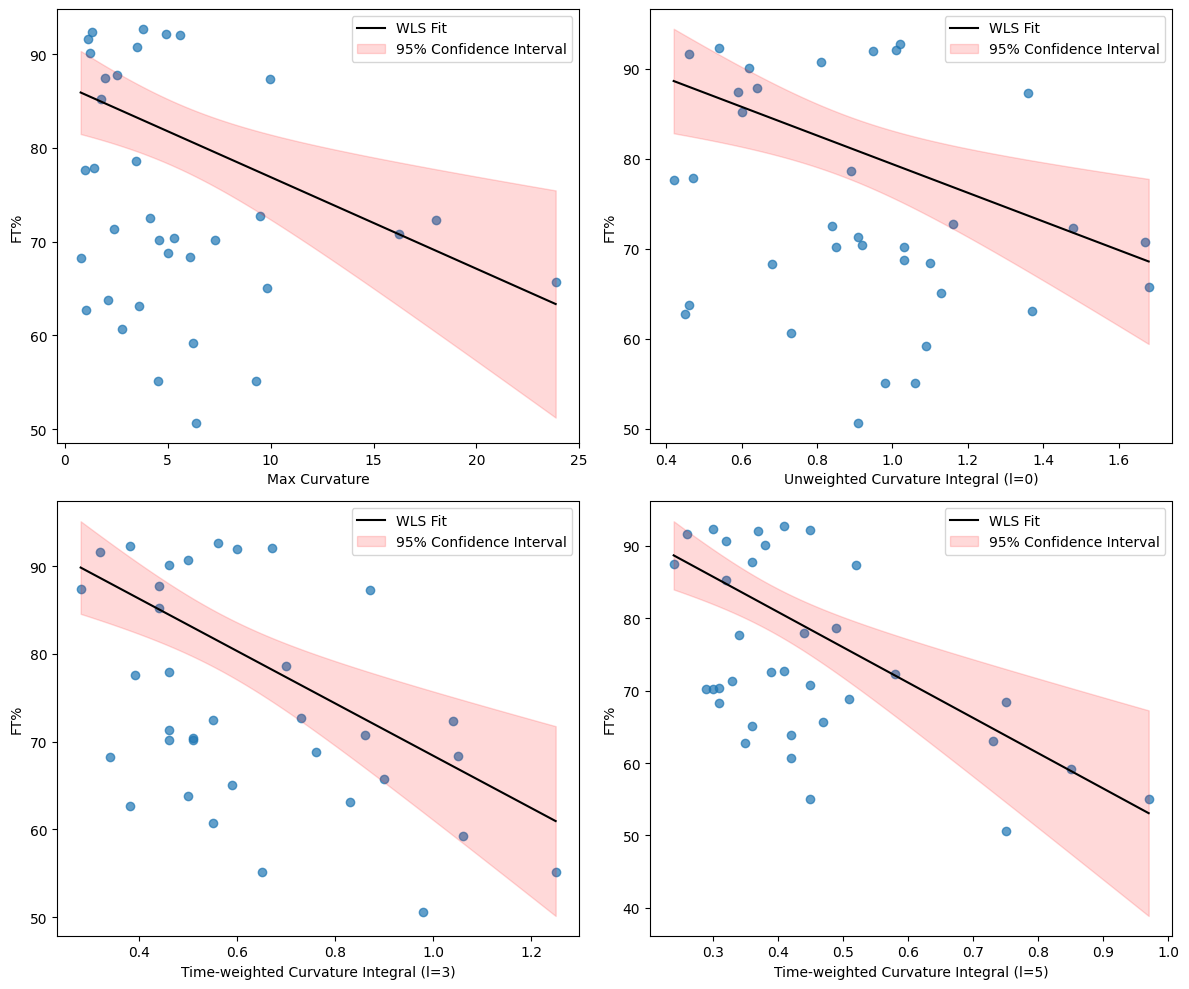}
        \caption{\it Weighted Least Squares (WLS) regression models illustrating the relationship between free throw percentage (FT\%) and four curvature-based shooting metrics. The black lines indicate the WLS regression fits, and the shaded red bands represent the 95\% confidence bands around each regression line. Across all subplots, higher curvature values are associated with lower FT\%.}
        \label{fig:regressionlines}
    \end{figure}

    Table \ref{tab:wls-results} summarizes the results of estimating the WLS model (\ref{eqn:weighted-least-squares}) for each of the four curvature metrics. Each metric is a statistically significant predictor of FT\%, with each $p$-value below 0.01. Among them, the time-weighted curvature integrals demonstrate the strongest explanatory power, particularly at $\ell = 5$, which achieves the highest $R^2$ value of 0.330. The confidence bands shown in Figure \ref{fig:regressionlines} illustrate this conclusion: the regression fits, particularly at $\ell = 3$ and $\ell = 5$, exhibit relatively more narrow and more stable bands, indicating greater precision in the estimated relationship between curvature and FT\% in these models. This finding aligns with the controlled experiments of \citet{slegers_role_2024}, suggesting that lower terminal curvature is correlated with greater shooting performance.

    \begin{table}[H]
        \centering
        \begin{tabular}{lrrr}
        \toprule
        \textbf{Metric} & \textbf{Coefficient} $\beta$ & $p$-\textbf{Value} & \textbf{R\textsuperscript{2}}\\
        \midrule
        Max Curvature                               &   --0.98  & 0.0043  & 0.222 \\
        Unweighted Curvature Integral ($\ell=0$)    &  --15.90  & 0.0053  & 0.213 \\
        Time-weighted Curvature Integral ($\ell=3$) &  --29.81  & 0.0004  & 0.316 \\
        Time-weighted Curvature Integral ($\ell=5$) &  --48.73  & 0.0003  & 0.330 \\
        \bottomrule
        \end{tabular}
        \caption{\it Summary statistics from the weighted least squares regression model (\ref{eqn:weighted-least-squares}) using each curvature metric to predict FT\%. The $p$-value tests the null hypothesis $\mathcal{H}_0 : \beta = 0$ against the one-sided alternative $\mathcal{H}_A : \beta < 0$.}
        \label{tab:wls-results}
    \end{table}
    
    Interestingly, the relationship between max curvature and shooting performance runs counter to the findings of \citet{slegers_role_2024}. We find that max curvature is negatively associated with FT\%. The present results reflect free throw shooting by NBA players in game conditions while the prior results reflect three-point jump shooting by lower-level professionals in practice conditions. Despite this discrepancy, the primary implications of our overall results are similar to those of past work. Placing greater emphasis on the latter part of the ball's trajectory---where control and release are most critical---improves the model's ability to explain variation in shooting accuracy.

\section{Discussion}

    \subsection{Results}
      
      The results of this study provide strong empirical support for the hypothesis that smoother shot trajectories---particularly those with less curvature near the point of release---are associated with greater free throw accuracy in real-game contexts in the NBA. Among the metrics examined, the time-weighted curvature integrals exhibited the strongest predictive power, particularly when using higher weighting parameters (\(\ell=3\) and \(\ell=5\)). This aligns with the intuition that late-stage control of the shot is critical for accuracy.
      
      Our findings are consistent in some ways with the conclusions drawn by \citet{slegers_role_2024}, and in some ways they differ. Both studies found that terminal ball path curvature is the strongest predictor of shooting performance and that coaches should focus on smoothing terminal curvature rather than smoothing max curvature. It is interesting to note that the two studies found directionally opposing conclusions regarding the relationship between max curvature and shooting performance. In the present study, max curvature is negatively associated with shooting performance.

      The primary difference between the studies is that the present used data from free throws attempted by NBA players in game conditions, and the former used data from three-point jump shots attempted by lower-level professionals in practice conditions. The contradictory results are especially interesting because from basketball domain knowledge, we would have expected lower max curvature to be more helpful for three-point shots. One-motion shots with lower max curvature allow the athlete to generate the greater power required for longer-distance shots. Explaining the observed phenomenon remains an open question. 
      
      Taken together, these findings contribute a more robust and generalizable framework for quantifying shot smoothness using in-game tracking data. They validate curvature-based metrics as scalable tools for performance evaluation, player development and automated scouting. Most importantly, they mark a shift from conventional, intuitive-driven coaching principles to data-informed validation with in-game evidence.

    \subsection{Limitations}

      While the curvature-based metrics developed in this study enhance the understanding of shooting mechanics using only ball tracking data, several limitations must be acknowledged. First, free throw shooting is a complex motion influenced by numerous factors beyond curvature, such as release angle, velocity and joint kinematics. Relying solely on ball tracking data and curvature analysis accounts for only a fraction of what determines shot accuracy, leaving many contributing factors unexplored. Second, the curvature metrics primarily assess the shot in the sagittal plane, neglecting lateral deviations that could introduce instabilities in the actual shot path. Third, Bézier curves are sensitive to parameters such as the number of control points and the selection of start and end points for the ball path. Since curvature-based metrics depend on the Bézier curves used to model the actual trajectory, small variations in these parameters can lead to differences in the computed metric values, which could influence the reproducibility of the model. Therefore, anyone using this approach should carefully choose control points and start/end criteria to best suit their specific needs and ensure consistency across the entire dataset.

      These findings might be valuable to coaches and players as quantifiable metrics and by helping them visualize the shooting trajectory, enhancing the understanding of free throw shooting mechanics and the identification of areas of improvements.

    \subsection{Future Work}
    
      Future work could involve integrating biomechanical data to expand the application of the developed metrics, enabling a more comprehensive analysis of how different joints coordinate to influence the smoothness of the shot. By employing a similar approach as the weighting function, the shooting path could be further segmented to isolate specific phases, allowing for targeted analyses based on different needs. Moreover, one could investigate the impact of release speed on shot smoothness, as it remains debatable and unclear.

  \bibliography{\ifnum\pdfstrcmp{\jobname}{output}=0 reports/free-throw-curvature\else ../free-throw-curvature\fi}

\begin{thebibliography}{13}
\expandafter\ifx\csname natexlab\endcsname\relax\def\natexlab#1{#1}\fi
\expandafter\ifx\csname url\endcsname\relax
  \def\url#1{{\tt #1}}\fi
\expandafter\ifx\csname urlprefix\endcsname\relax\def\urlprefix{URL }\fi

\bibitem[{{Basketball Reference}(2024)}]{basketball_reference_2023-24_2024}
{Basketball Reference} (2024).
\newblock 2023-24 {NBA} {Player} {Stats}: {Totals}.
\newline\urlprefix\url{https://www.basketball-reference.com/leagues/NBA\_2024\_totals.html}

\bibitem[{Baydas \& Karakas(2019)}]{baydas_defining_2019}
Baydas, S., \& Karakas, B. (2019).
\newblock Defining a curve as a {Bezier} curve.
\newblock {\em Journal of Taibah University for Science\/}, {\em 13\/}(1), 522--528.

\bibitem[{Button et~al.(2003)Button, Macleod, Sanders, \& Coleman}]{button_examining_2003}
Button, C., Macleod, M., Sanders, R., \& Coleman, S. (2003).
\newblock Examining movement variability in the basketball free-throw action at different skill levels.
\newblock {\em Research Quarterly for Exercise and Sport\/}, {\em 74\/}(3), 257--269.

\bibitem[{Haefner(2010)}]{haefner_7_2010}
Haefner, J. (2010).
\newblock 7 {Tips} {To} {Improve} {Your} {Shooting} {Mechanics}.
\newline\urlprefix\url{https://www.usab.com/news/2014/01/7-tips-to-improve-your-shooting-mechanics}

\bibitem[{Hamilton \& Reinschmidt(1997)}]{hamilton_optimal_1997}
Hamilton, G.~R., \& Reinschmidt, C. (1997).
\newblock Optimal trajectory for the basketball free throw.
\newblock {\em Journal of Sports Sciences\/}, {\em 15\/}(5), 491--504.

\bibitem[{Kozar et~al.(1994)Kozar, Vaughn, Whitfield, Lord, \& Dye}]{kozar_importance_1994}
Kozar, B., Vaughn, R.~E., Whitfield, K.~E., Lord, R.~H., \& Dye, B. (1994).
\newblock Importance of {Free}-{Throws} at {Various} {Stages} of {Basketball} {Games}.
\newblock {\em Perceptual and Motor Skills\/}, {\em 78\/}(1), 243--248.

\bibitem[{Mullineaux \& Uhl(2010)}]{mullineaux_coordination-variability_2010}
Mullineaux, D.~R., \& Uhl, T.~L. (2010).
\newblock Coordination-variability and kinematics of misses versus swishes of basketball free throws.
\newblock {\em Journal of Sports Sciences\/}, {\em 28\/}(9), 1017--1024.

\bibitem[{{National Basketball Association}(2016)}]{national_basketball_association_nba_2016}
{National Basketball Association} (2016).
\newblock {NBA} announces multiyear partnership with {Sportradar} and {Second} {Spectrum}.
\newline\urlprefix\url{https://pr.nba.com/nba-announces-multiyear-partnership-sportradar-second-spectrum/}

\bibitem[{Pakosz et~al.(2021)Pakosz, Domaszewski, Konieczny, \& B\k{a}czkowicz}]{pakosz_muscle_2021}
Pakosz, P., Domaszewski, P., Konieczny, M., \& B\k{a}czkowicz, D. (2021).
\newblock Muscle activation time and free-throw effectiveness in basketball.
\newblock {\em Scientific Reports\/}, {\em 11\/}, 7489.

\bibitem[{Penny(2016)}]{penny_overlooked_2016}
Penny, R. (2016).
\newblock The {Overlooked} {Importance} of {Arm} \& {Wrist} {Angles}.
\newline\urlprefix\url{https://www.breakthroughbasketball.com/fundamentals/shooting-arm-wrist-angle.html}

\bibitem[{Slegers(2022)}]{slegers_basketball_2022}
Slegers, N. (2022).
\newblock Basketball shooting performance is maximized by individual-specific optimal release strategies.
\newblock {\em International Journal of Performance Analysis in Sport\/}, {\em 22\/}(3), 393--406.

\bibitem[{Slegers \& Love(2024)}]{slegers_role_2024}
Slegers, N., \& Love, D. (2024).
\newblock The role of ball path curvature in basketball shooting accuracy.
\newblock {\em Journal of Sports Sciences\/}, {\em 42\/}(21), 2052--2060.

\bibitem[{Tran \& Silverberg(2008)}]{tran_optimal_2008}
Tran, C.~M., \& Silverberg, L.~M. (2008).
\newblock Optimal release conditions for the free throw in men's basketball.
\newblock {\em Journal of Sports Sciences\/}, {\em 26\/}(11), 1147--1155.

\end{thebibliography}

\end{document}